\documentclass[prd,aps]{revtex4}
\usepackage{psfig}
\begin{document}
\title{Detector Depth Dependence of the High Energy Atmospheric Neutrino Flux}
\date{\today}
\author{John M.~LoSecco}
\email{losecco@undhep.hep.nd.edu}
\affiliation{Physics Department, University of Notre Dame, Notre Dame, Indiana
46556 USA}
\begin{abstract}
We note that detector depth can influence the decay path length available for
the primary and secondary particles that are the source of atmospheric
neutrinos.  As a consequence there is a location dependent
modulation to the neutrino flux, which could be as large as 5-10\% in some
directions.
\end{abstract}

\maketitle

\section{Introduction}
\begin{figure}
\mbox{\psfig{file=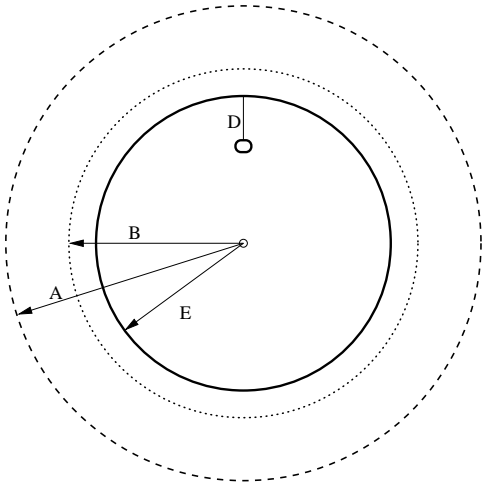,width=2.5in}}
\caption{\label{geom}The atmospheric neutrino source geometry.  This figure
is not to scale}
\end{figure}
Estimates for the ambient neutrino flux are an important input parameter
to experiments analyzing underground neutrino interactions and upward
going muons to understand the time evolution of a neutrino beam.
Different directions provide neutrino fluxes with varying source distance
so observation of the angular distribution is an essential tool in the
study of neutrino oscillations.
Since the overall flux normalization is uncertain, experiments frequently
place a greater emphasis on the shape of the distribution than the absolute
event rate.

This note points out {\em depth dependent} effects that can also provide
a directional modulation.  These effects are modest, but predominantly effect
the higher end of the neutrino spectrum.

Upward going muons \cite{macro,kam,imb,mcgrath,oyama}
are usually attributed to muon neutrino interactions
in the rock surrounding the detector.
In order to be detected as an upward going muon, the lepton produced in
a neutrino interaction must propagate through the rock to be recorded in
the detector.  If we approximate the muon energy loss as muon energy
independent then the range increases linearly with the muon energy.  So
the target rock surrounding the detector has a larger effective mass
for neutrino interactions at high
energy, scaling roughly as $E_{\nu}^2$.  Over a substantial range of
neutrino energies the cross section rises linearly.  So that a constant mass
detector will have more high energy neutrino interactions than low energy
neutrino interactions at the same flux.  These two factors suggest that
the neutrino induced muon flux is sensitive to the third power of the
neutrino energy.  Small neutrino flux differences at high energies are
substantially amplified by this $E_{\nu}^3$ factor.

We present a one dimensional model to show that the atmospheric decay path
length is a function of the detector depth.  Detectors which are above sea
level will see neutrinos with a higher decay path length than detectors
below sea level.  To first order the high energy part of the neutrino flux
is proportional to the decay path length.

\section{Geometry}
\begin{figure}
\mbox{\psfig{file=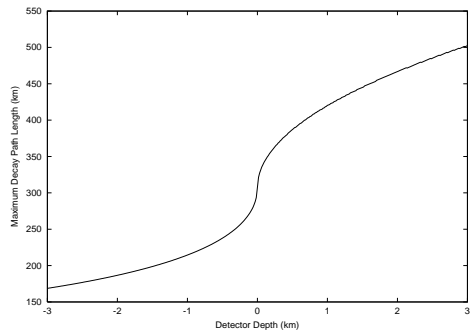,width=2.5in}}
\caption{\label{maxrel}The maximum decay path length as a function of detector
depth.  The curve is the maximum decay length relative to the maximum at zero
detector depth.  The depth axis runs from 3 km below to 3 km above
the surface,}
\end{figure}
Figure \ref{geom} illustrates the geometry.  This figure is not to scale.
We take $E$ to be the radius of the Earth, 6380 km and $A$ to be the radius
at which neutrino source particles are produced.  $B$ will be one interaction
length below $A$.  Most decays will occur between $A$ and $B$.
$D$ represents the detector depth.  If the detector is above sea level
$D$ will be negative.  We take as the decay length the difference in length
for ray originating at $D$ and ending at a point along the ray at radius
$A$ or $E$.  It should be clear from figure \ref{geom}, with its
disproportionate scale that the decay length will depend on detector depth.
For muons we take $B=E$, the surface of the Earth.

\begin{figure}
\mbox{\psfig{file=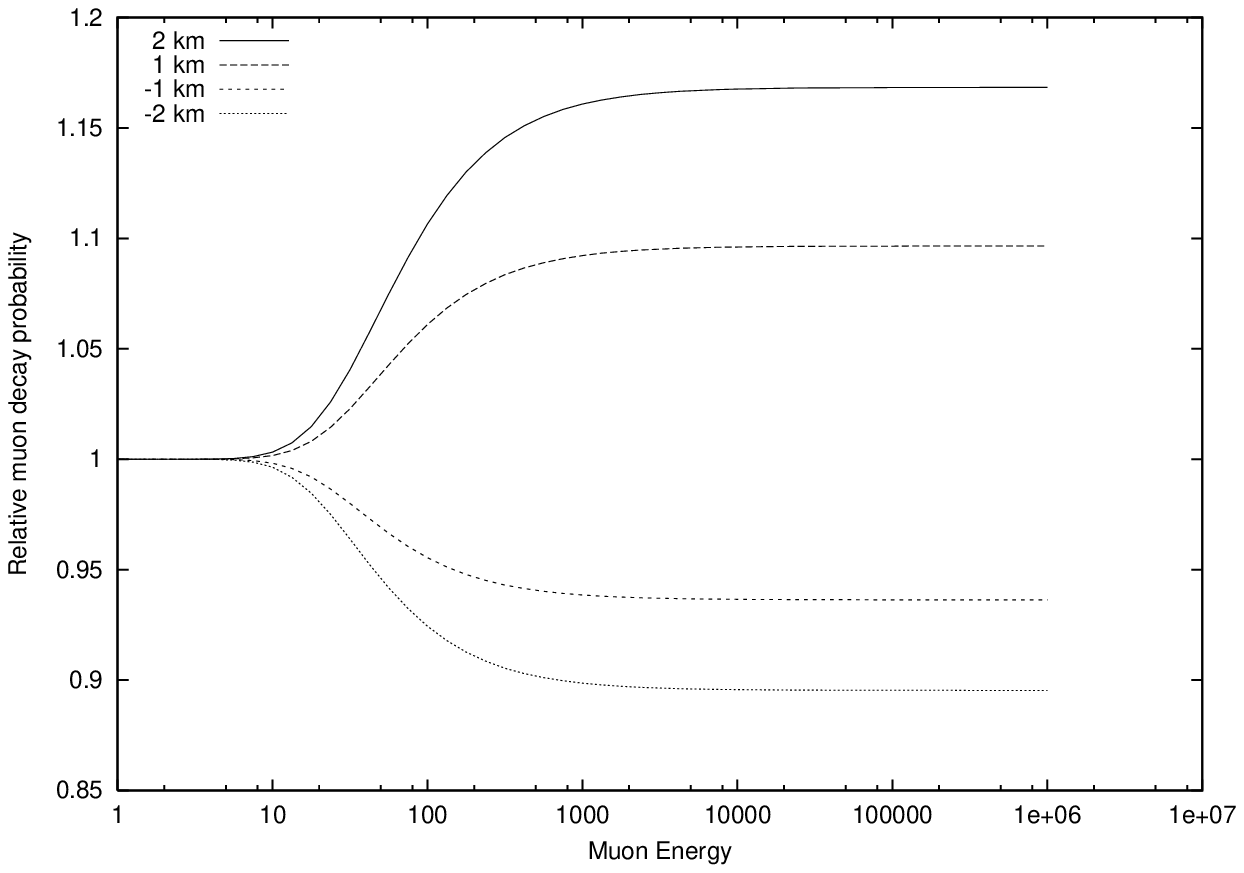,width=2.2in} \psfig{file=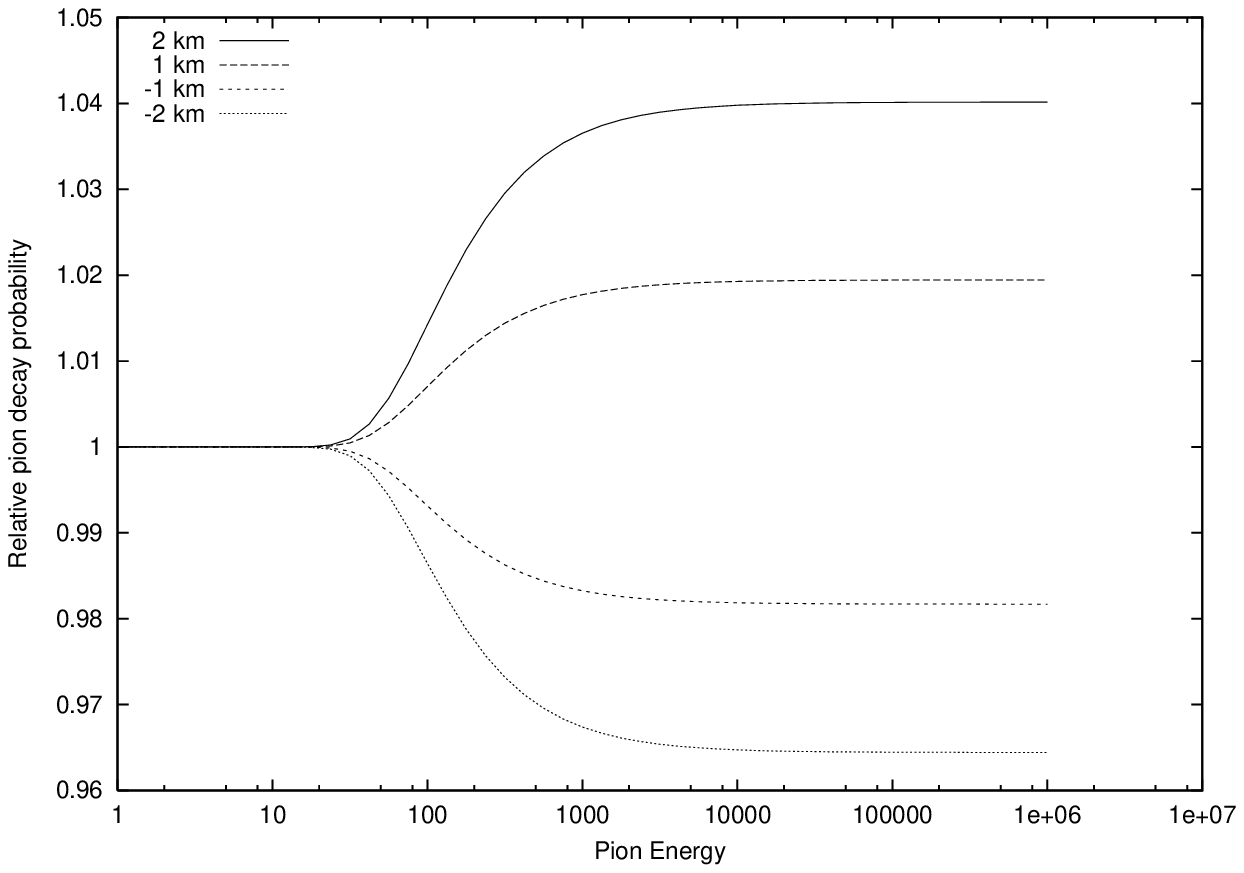,width=2.2in}
\psfig{file=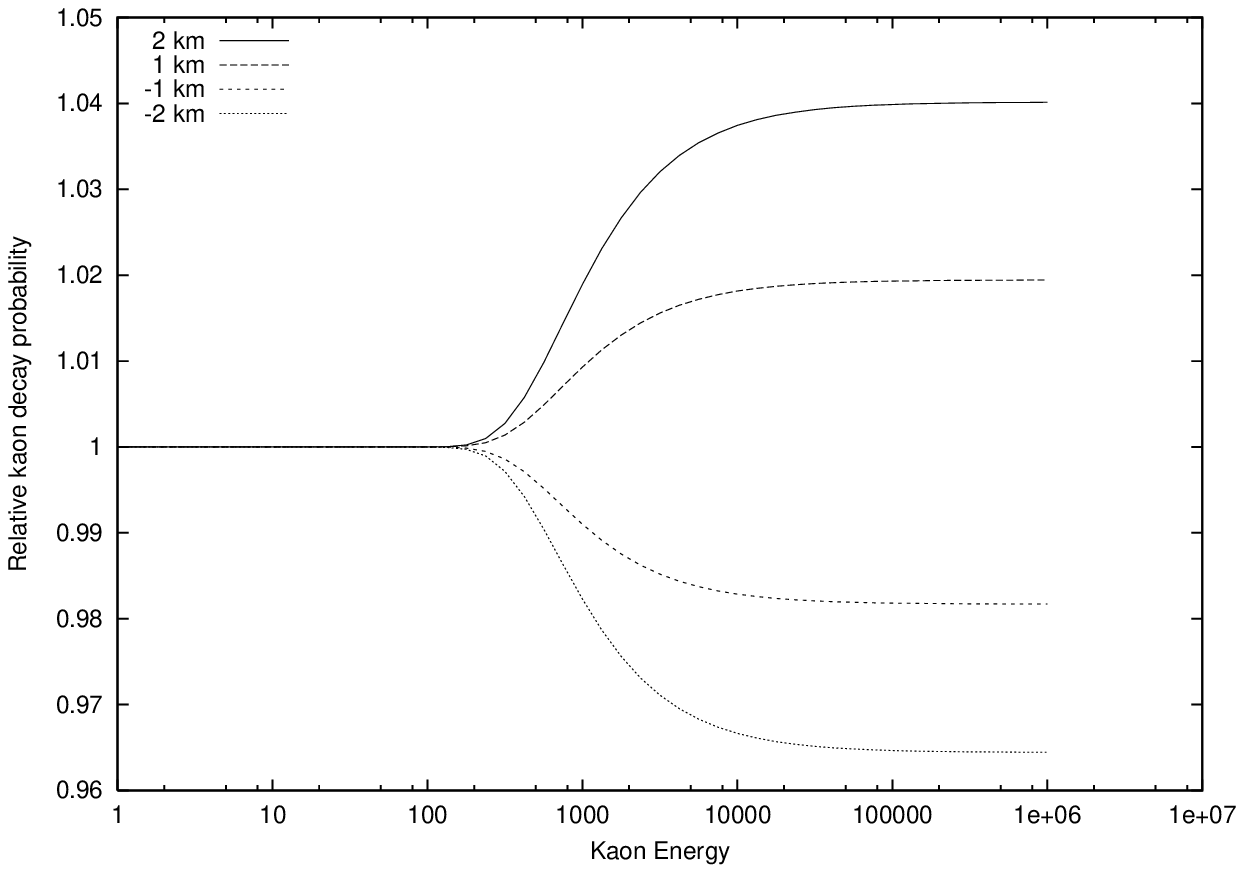,width=2.2in}}
\caption{\label{murel}The relative muon, pion and kaon decay probability as a
function of
energy.  The four curves are for four different detector depths, -2, -1, 1
and 2 km.
This is averaged over the zenith angle bin $-0.1< \cos(\theta_{z}) <0. $}
\end{figure}

A particle initiating in the upper atmosphere will travel a distance S before
being absorbed.
\[
S=
\sqrt{A^{2}-\sin{\theta_{z}}^{2} (E-D)^{2}}
-\sqrt{B^{2}-\sin{\theta_{z}}^{2} (E-D)^{2}}
\]
Very near the horizon, the particle path may not intersect $B$, if the
detector is above sea level ($\sin{\theta_{z}}^{2} (E-D)^{2} > B^{2}$).
In that case we approximate the decay length
by the distance from the upper atmosphere to the detector.
\[
S=-(E-D) \cos{\theta_{z}} + \sqrt{A^{2}-\sin{\theta_{z}}^{2} (E-D)^{2}}
\]
$\theta_{z}$ is the zenith angle.
Note for upward going neutrinos $\cos{\theta_{z}} < 0$.

In figure \ref{maxrel} we illustrate this effect for muons, where we take $B=E$
and $A=E+7.4$ km.  The figure shows
the maximum decay length for muons below the horizon.  In most cases this
maximum is obtained at the horizon.  But when the detector is above sea level,
$D<0$, the maximum decay path length is achieved near the horizon.  Notice that
the decay path length increases with the detector height.

\begin{figure}
\mbox{\psfig{file=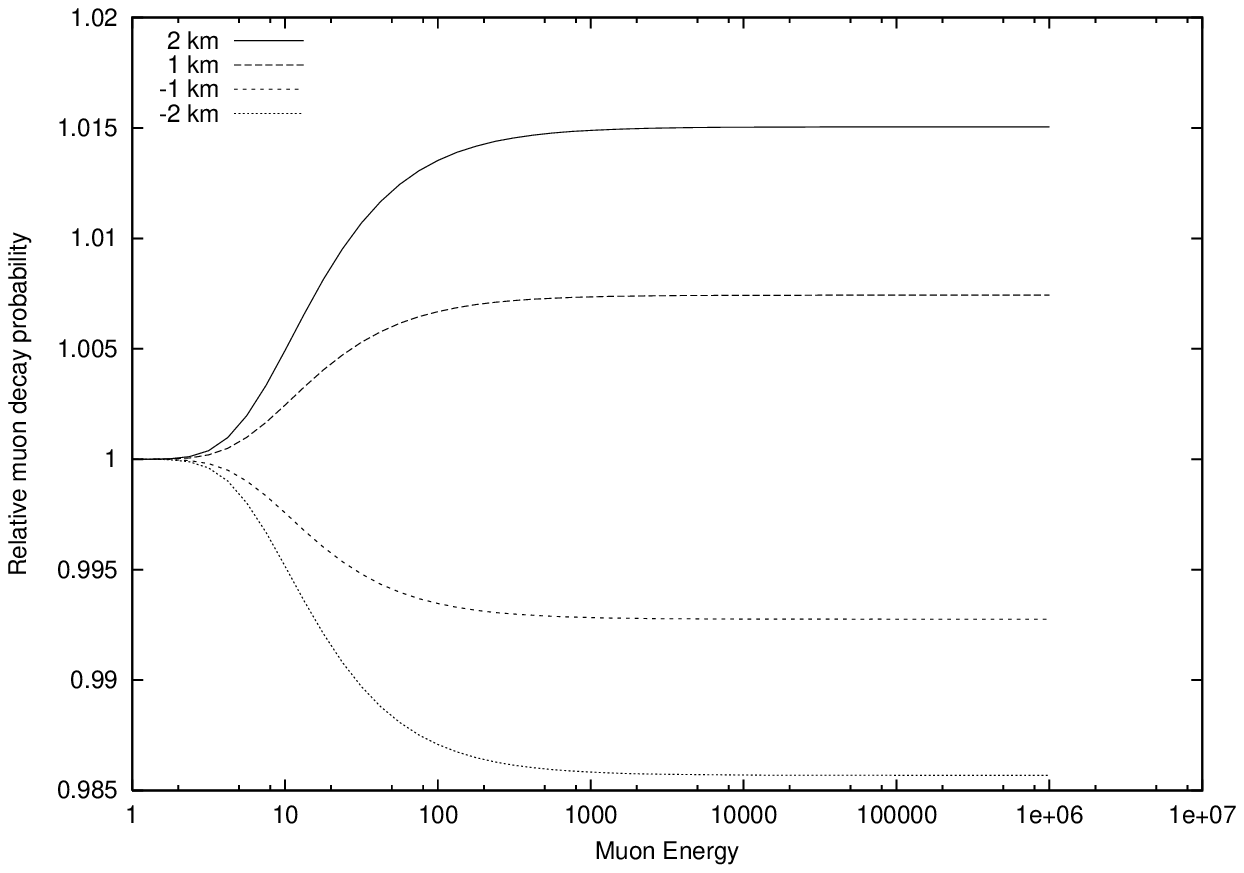,width=2.2in}
\psfig{file=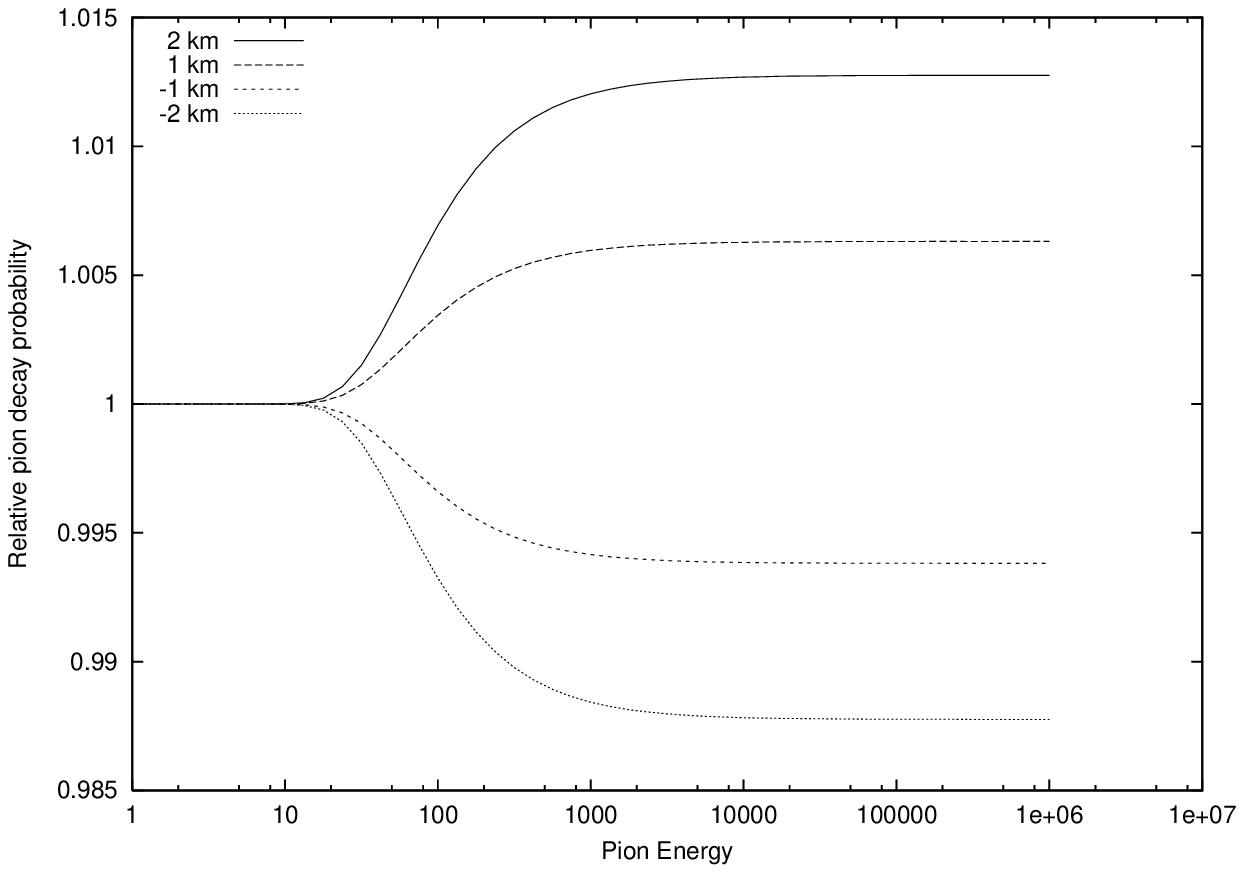,width=2.2in}
\psfig{file=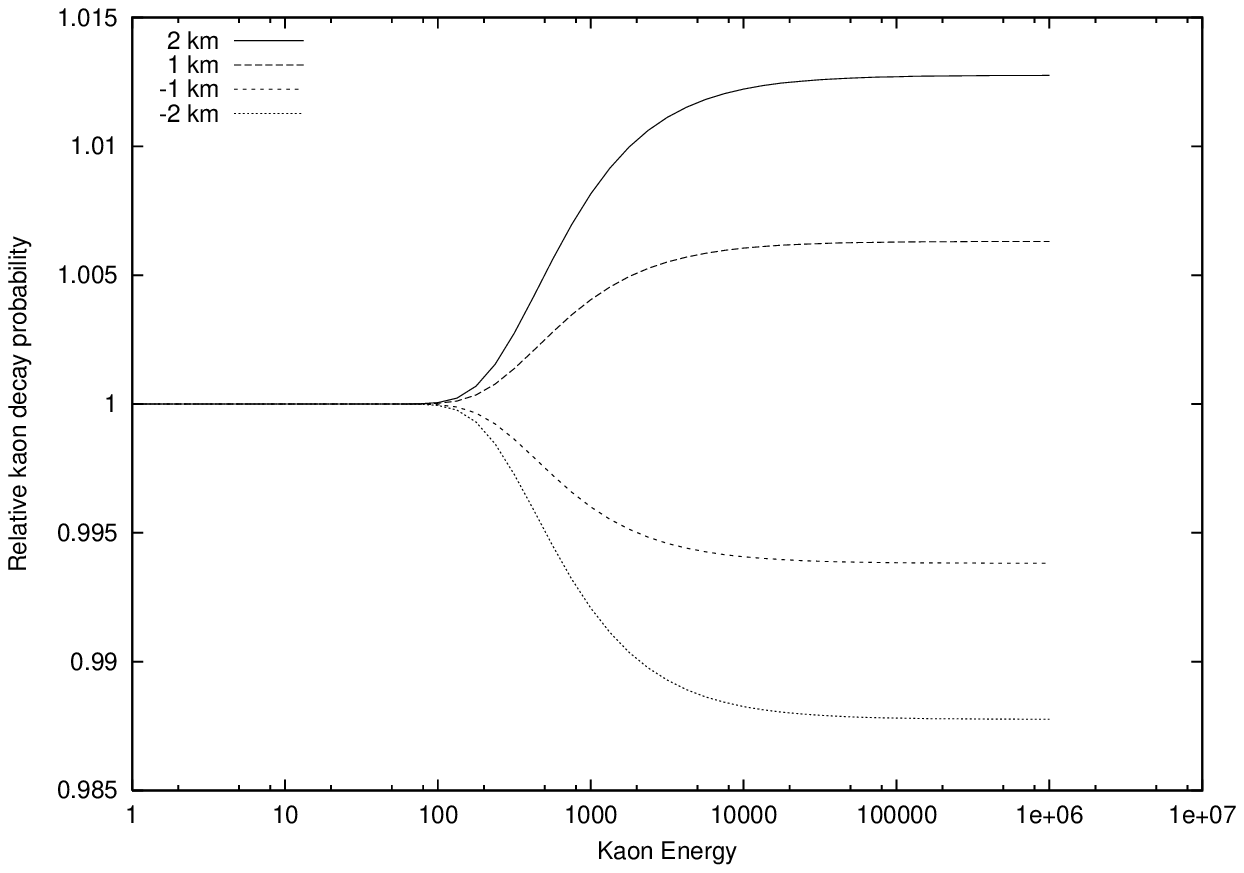,width=2.2in}}
\caption{\label{murel2}The relative muon, pion and kaon decay probability as a
function of
energy.  The four curves are for four different detector depths.
This is averaged over the zenith angle bin $-0.2< \cos(\theta_{z}) <-0.1 $}
\end{figure}

Common detector depths, $D$ are in the range of -1 km $<D<$ 2 km.
Detectors located in the mountains tend to be deep underground but well
above sea level.

\section{Magnitude of the Effect}
In figure \ref{murel} we explore the relative flux, as a function of neutrino
parent particle energy, for 4 different detector depths.  This figure plots the
contribution to the neutrino flux of a detector at $D=$-2, -1, 1 or 2 km.
relative to the contribution to the neutrino
flux for a detector located at sea level ($D=0$).  The flux is averaged over
the solid angle region of $-0.1< \cos(\theta_{z}) <0. $, the angular  bin
just below the horizon.  Variations are about 5\% to 10\%.

\begin{figure}
\mbox{\psfig{file=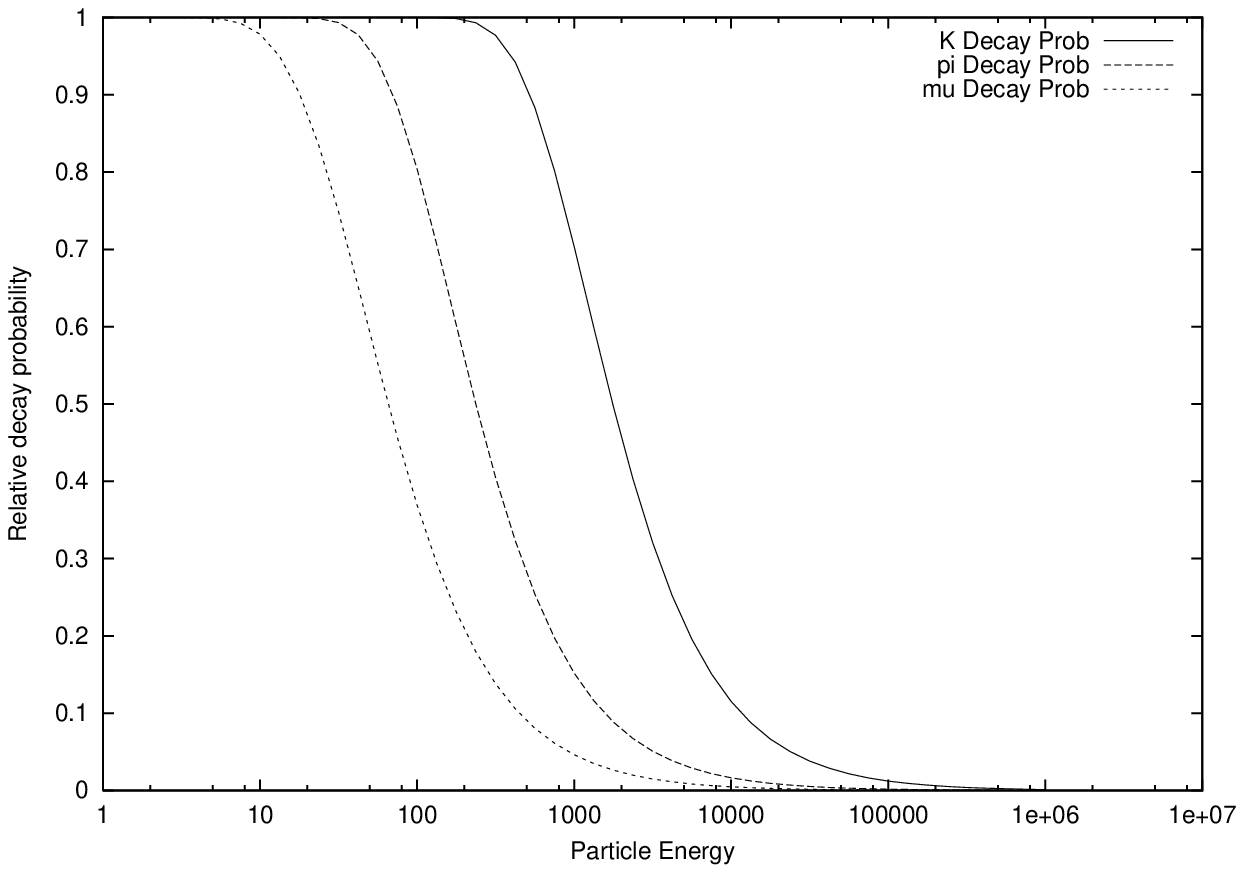,width=2.5in}
\psfig{file=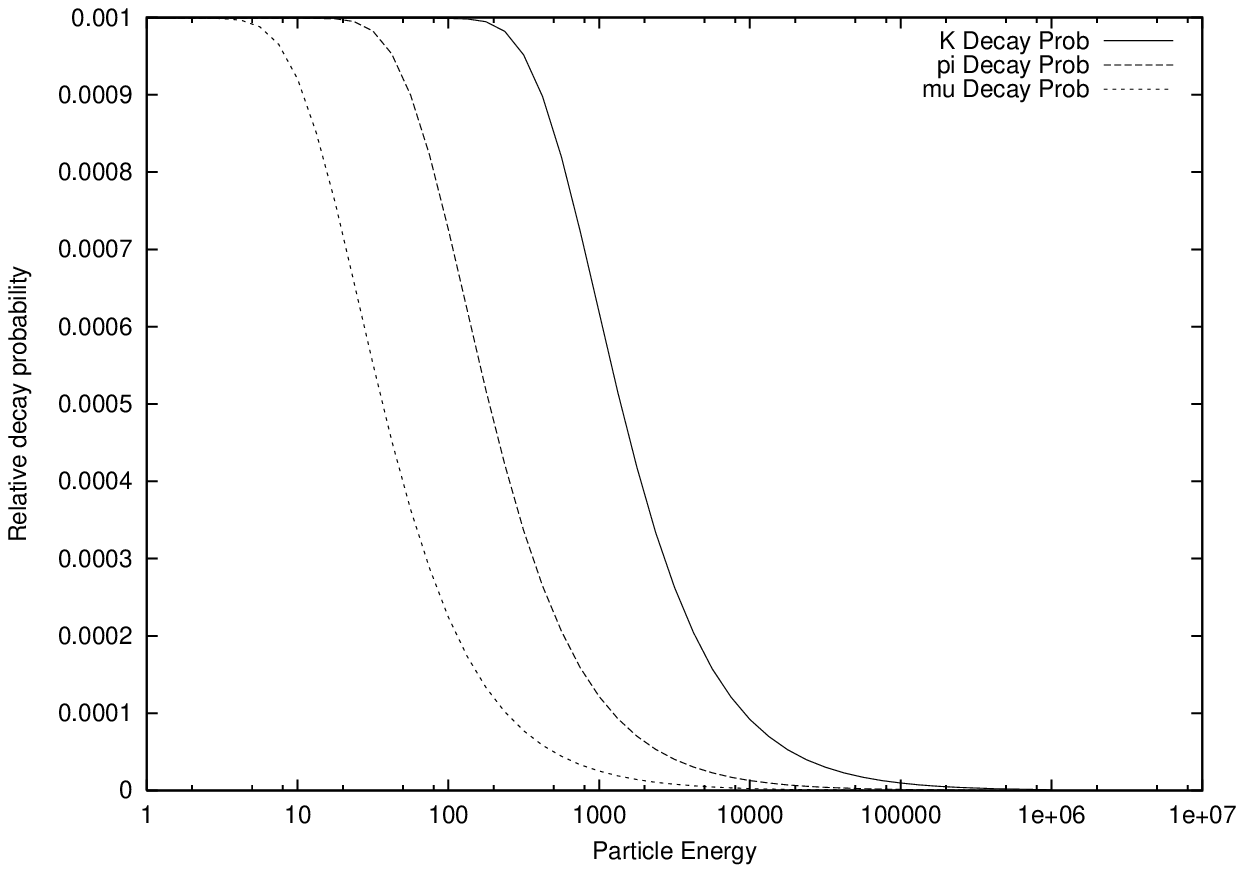,width=2.5in}}
\caption{\label{mpkprob}The muon, pion and kaon decay probability as a
function of energy for $D=0$.
The one on the left is averaged over the zenith angle bin
$-0.1< \cos(\theta_{z}) <0. $.  The one on the right is averaged over
$-0.2< \cos(\theta_{z}) <-0.1 $.  The shorter path lengths in the
$-0.2< \cos(\theta_{z}) <-0.1 $ region yield a softer flux and lower event
rate}
\end{figure}
In figure \ref{murel} one sees that the enhancement is not present at low
energies, where all particles will decay.  There is a transition region at
moderate energies where the decay length is comparable to the absorption
length.  At the highest energies the decay probability scales linearly with
the available decay length so the flux differences directly reflect the
path length differences due to detector depth.
The shape differences for
muons, pions and kaons are due to the differences in masses, lifetimes and
absorption lengths.

Figure \ref{murel2} is similar to figure \ref{murel} except that now the solid
angle region $-0.2< \cos(\theta_{z}) <-0.1 $ is considered.
Variations are now of the order of 0.6\% to 1.2\%.  The much lower flux
modification away from the horizon indicates that standard neutrino flux
calculations, that do not include the detector depth, will not correctly
represent the angular distribution.

Figure \ref{mpkprob} plots the muon, pion and kaon decay probability as a
function of decay particle energy.  The contribution at high energies is
dominated by the particle with the shortest lifetime.

\begin{figure}
\mbox{\psfig{file=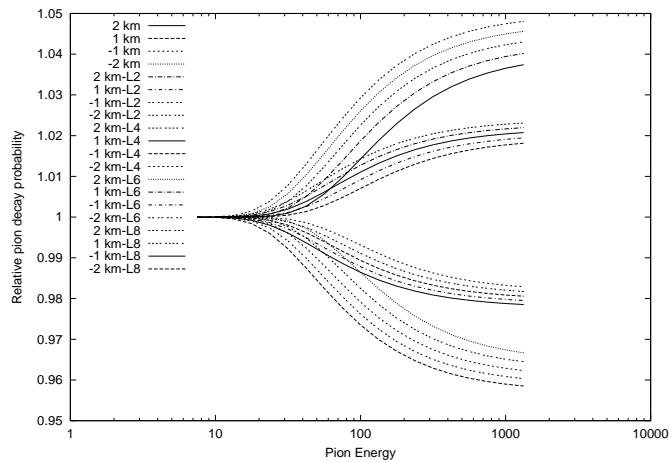,width=3.5in}}
\caption{\label{shower}Hadronic shower development brings the enhancement to
lower energies.  The 4 branches are the 4 detector depths considered in figures
\ref{murel} and \ref{murel2}.  The multiple curves in each branch are
for hadrons initiating 2 absorption lengths lower in the atmosphere than for
the curves to their right.}
\end{figure}

Figure \ref{shower} illustrates the depth effect in hadronic shower evolution.
The multiple curves are for hadrons initiating at greater depth into the
atmosphere.  Each curve is 2 absorption lengths deeper in than the one to
its right.  The depth modulation is maintained at approximately the same
amplitude, but as showers go deeper into the atmosphere the curves move to
lower energies since the absorption length drops.  Higher energy hadrons are
more likely to be absorbed than to decay when the are formed lower in the
atmosphere.

\section{Limitations}
This paper has made no attempt to quantitatively sum the various contributions
to the atmospheric neutrino flux.  Muons, pions and kaons all contribute but
their relative contributions depend on their initial fractions and the fate of
the other contributions.  At very high energies ``prompt'' sources of neutrinos
such as charm decay become important.  At the highest energies muons produced
via decays of the form $\rho \rightarrow \mu^+ \mu^-$ and
$J/\psi \rightarrow \mu^+ \mu^-$, or Drell-Yan processes constitute a
significant source of neutrino flux.

\section{Conclusions}
Our analysis suggests that there is a modest location dependent contribution
to the high energy atmospheric neutrino flux.  Detectors above sea level
(but still underground) will see enhancements of the high energy flux in
the vicinity of the horizon.  This would be manifest as an angular distortion
and an increased rate.  Detectors below sea level would expect to see the
opposite effect, with fewer high energy events and fewer events near the
horizon.  While these effects are of the order of 5\% to 10\% they contribute
a systematic distortion when comparing data taken at different locations or
when comparing observations from a deep detector to a flux estimate for a
sea level location.
 
\section{Acknowledgments}
Todor Stanev has recently reached similar conclusions
about the influence of detector depth.
I would like to thank O.~Ryazhskaya for pointing out the significance of
muon production via hadronic resonance decay ($\rho \rightarrow \mu^+ \mu^-$)
as a major source of neutrinos at high energies.
This work was supported in part by the US Department of Energy
under grant DE-FG02-00ER41145.

\end{document}